\documentclass[a4paper, 12pt]{article}

\usepackage{amsmath, amssymb, amsthm}
\usepackage[english]{babel}
\usepackage[T1]{fontenc}

\newcommand{\be}{\begin{equation}}
\newcommand{\ee}{\end{equation}}
\newcommand{\bd}{\begin{displaymath}}
\newcommand{\ed}{\end{displaymath}}
\newcommand{\ba}{\begin{eqnarray}}
\newcommand{\ea}{\end{eqnarray}}

\def\R{{I \!\! R}}

\def\a{\alpha}

\def\v12{(v-w)}

\def\({\left(}
\def\){\right)}

\def\bgr#1\egr{{\allowdisplaybreaks\begin{gather}#1\end{gather}}}
\def\bma#1\ema{{\allowdisplaybreaks\begin{align}#1\end{align}}}

\def\oplem#1{\begin{lemma}\, {\rm #1}\, \it }
\def\cllem{\end{lemma}\rm \par }
\def\opthm#1{\begin{theorem}\, {\rm #1}\, \it }
\def\clthm{\end{theorem}\rm \par }

\def\R{\mathbb{R}}

\newcommand{\fer}[1]{(\ref{#1})}
\newcommand{\bq}{\begin{equation}}
\newcommand{\eq}{\end{equation}}
\def\bqa{\begin{eqnarray}}
\def\eqa{\end{eqnarray}}
\def\bd{\begin{displaymath}}
\def\ed{\end{displaymath}}

\newtheorem{thm}{Theorem}

\theoremstyle{remark}

\theoremstyle{definition}

\newenvironment{equations}{\equation\aligned}{\endaligned\endequation}

\begin{document}

\title{A concavity property for the reciprocal of Fisher information and its consequences on Costa's EPI}

 \author{Giuseppe Toscani \thanks{Department of Mathematics, University of Pavia, via Ferrata 1, 27100 Pavia, Italy.
\texttt{giuseppe.toscani@unipv.it} }}

\maketitle

\begin{center}\small
\parbox{0.85\textwidth}{
\textbf{Abstract.} We  prove that the reciprocal of Fisher information of a log-concave
probability density $X$ in $\R^n$ is concave in $t$ with respect to the addition of a
Gaussian noise $Z_t = N(0, tI_n)$. As a byproduct of this result we show that the
third derivative of the entropy power of a log-concave probability density $X$ in
$\R^n$ is nonnegative in $t$ with respect to the addition of a Gaussian noise $Z_t$.
For log-concave densities this improves the well-known Costa's concavity pro\-per\-ty
of the entropy power \cite{Cos}.

}
\end{center}

\medskip

\section{Introduction}

Given a random vector $X$ in $\R^n$, $n \ge 1$ with density $f(x)$, let
 \be\label{Shan}
H(X) = H(f) = - \int_{\R^n} f(x) \log f(x)\, dx
 \ee
denote its entropy functional (or Shannon's entropy). The entropy power introduced by
Shannon \cite{Sha} is defined by
  \be\label{ep}
 N(X) = N(f) = \frac1{2\pi e}\exp\left(\frac 2n H(X)\right).
 \ee
The entropy power is built to be linear at  Gaussian random vectors. Indeed, let $Z_\sigma = N(0,\sigma I_n)$ denote the $n$-dimensional Gaussian random vector having mean vector $0$ and covariance matrix $\sigma I_n$, where $I_n$ is the identity matrix. Then $N(Z_\sigma) = \sigma$. Shannon's entropy power inequality (EPI), due to Shannon and Stam \cite{Sha, Sta} (cf. also
\cite{Cos, GSV, GSV2, Rio, Tos3, ZF} for other proofs and extensions) gives a lower bound on
Shannon's entropy power of the sum of independent random variables $X, Y$ in $\R^n$ with densities
 \be\label{entr}
N(X+Y) \ge N(X) + N(Y),
 \ee
with equality if and only $X$ and $Y$ are Gaussian random vectors with proportional
covariance matrices.

In $1985$ Costa \cite{Cos} proposed a stronger version of EPI \fer{entr}, valid for the
case in which $Y = Z_t$, a Gaussian random vector independent of $X$.
In this case
 \be\label{strong}
N(X+Z_t) \ge (1-t)N(X) + t N(X+ Z_1), \qquad 0\le t \le 1
 \ee
or, equivalently, $N(X+Z_t)$,  is concave in $t$, i.e.
 \be\label{conc}
\frac{d^2}{dt^2} N(X+Z_t) \le 0.
 \ee
Note that equality to zero in \fer{conc} holds if and only if $X$ is a Gaussian random
variable, $X= N(0, \sigma I_n)$. In this case, considering that $Z_\sigma$ and $Z_t$ are independent each other, and Gaussian densities are
stable under convolution,
$N(Z_\sigma + Z_t) = N(Z_{\sigma+t}) = \sigma +t$, which implies
  \be\label{conc=0}
\frac{d^2}{dt^2} N(Z_\sigma +Z_t) = 0.
 \ee
Let now consider, for a given  random vector $X$ in $\R^n$ with smooth density,  its
Fisher information
 \be\label{fish}
I(X) = I(f) = \int_{\{f>0\}} \frac{|\nabla f(x)|^2}{f(x)} \, dx.
 \ee
Blachman--Stam inequality \cite{Bla, Dem, Sta} gives a lower  bound on the reciprocal of Fisher information of the sum of
independent random vectors with (smooth) densities
 \be\label{BS}
\frac 1{I(X+Y)} \ge \frac 1{I(X)}+ \frac 1{I(Y)},
 \ee
still with equality if and only $X$ and $Y$ are Gaussian random vectors with
proportional covariance matrices. 

In analogy with the definition of entropy power, let us introduce the (normalized) reciprocal of Fisher information
 \be\label{ent-fish}
 \tilde I(X) = \frac n{I(X)}.
 \ee
By construction, since $I(Z_\sigma) = n/\sigma$,  $\tilde I(\cdot)$ is linear at
Gaussian random vectors, with $\tilde I(Z_\sigma)= \sigma$. Moreover, in terms of
$\tilde I$, Blachman--Stam inequality reads
 \be\label{BSn}
 \tilde I(X+Y) \ge  \tilde I(X) + \tilde I(Y).
 \ee
Therefore, both the entropy power \fer{ep} and the reciprocal of Fisher information $\tilde I$, as given by \fer{ent-fish}, share
common properties when evaluated on Gaussian random vectors and on sums of independent random vectors.

By pushing further this analogy, in agreement with Costa's result on entropy power, we
will prove that the quantity $\tilde I(X+Z_t)$ satisfies the analogous of inequality
\fer{strong}, i.e.
 \be\label{strong2}
 \tilde I(X+Z_t) \ge (1-t)\tilde I(X) + t \tilde I(X+ Z_1), \qquad
0\le t \le 1
 \ee
or, equivalently
 \be\label{conc2}
\frac{d^2}{dt^2} \tilde I(X+Z_t) \le 0.
 \ee
Unlike Costa's result, the proof of \fer{conc2} is restricted to log-concave random vectors. Similarly to \fer{conc},  
equality to zero in \fer{conc2} holds if and only if $X$ is a Gaussian random vector,
$X= N(0, \sigma I_n)$.

The estimates obtained in the proof of  \fer{conc2} can be fruitfully employed to study the third derivative of $N(X+Z_t)$. The surprising result is that, at least for log-concave probability densities, the third derivative has a sign, and
 \be\label{third}
\frac{d^3}{dt^3} N(X+Z_t) \ge 0.
 \ee
Once again, equality to zero in \fer{third} holds if and only if $X$ is a Gaussian random
variable, $X= N(0, \sigma I_n)$.
Considering that
 \[
\frac{d}{dt} N(X+Z_t) \ge 0,
 \]
the new inequality \fer{third} seems to indicate that the subsequent derivatives of $N(X+Z_t)$ alternate in sign, even if a proof of this seems prohibitive.

The concavity property of the reciprocal of Fisher information is a consequence of a recent result of the present author \cite{Tos4} related to the functional
 \begin{equations}\label{j2}
 J(X)= J(f) = \sum_{i,j=1}^n \int_{\{f>0\}} \left[\partial_{ij}(\log f) \right]^2f \, dx = \\
\sum_{i,j=1}^n \int_{\{ f>0\}} \left[\frac{\partial_{ij}f}f -
\frac{\partial_{i}f\partial_{j}f}{f^2} \right]^2 f \, dx.
 \end{equations}
We remark that, given a random vector $X$ in $\R^n$, $n \ge 1$, the functional $J(X)$
is well-defined for a smooth, rapidly decaying probability density $f(x)$ such that
$\log f$ has growth at most polynomial at infinity. As proven by Villani in
\cite{Vil}, $J(X)$ is related to Fisher information by the relationship
 \be\label{JJ}
 J(X+ Z_t) = - \frac d{dt}I(X+Z_t).
 \ee
 The main result in \cite{Tos4} is a new inequality for $J(X+Y)$, where
$X$ and $Y$ are independent random vectors in $\R^n$, such that their probability
densities $f$ and $g$ are \emph{log-concave}, and $J(X)$, $J(Y)$ are well defined. For
any constant $\alpha$, with $0 \le \alpha \le 1$, it holds
 \begin{equation}\label{ine-main}
 J(X+Y) \le \alpha^4 J(X) + (1-\alpha)^4J(Y) + 2\alpha^2(1-\alpha)^2 H(X,Y),
 \end{equation}
where
 \be\label{rest}
 H(X,Y) = \sum_{i,j=1}^n \int_{\{f>0\}} \frac{\partial_i f\partial_j f}f \, dx
 \int_{\{g>0\}} \frac{\partial_i g\partial_j g}g \, dx.
 \ee
Note that, in one-dimension $H(f,g) = I(f)I(g)$. Inequality \fer{ine-main} is sharp.
Indeed, there is equality if and only if $X$ and $Y$ are $n$-dimensional Gaussian
vectors with covariance matrices proportional to $\alpha I_n$ and $(1-\alpha)I_n$
respectively.

Even if inequality \fer{ine-main} is restricted to the set of log-concave densities,
this set  includes many of the most commonly-encountered parametric families of
probability density functions \cite{MO}.

Inequality \fer{ine-main} implies a Blachman-Stam type inequality for $\sqrt{J(\cdot)}$ \cite{Tos4}
  \be\label{sharp2}
\frac 1{\sqrt{J(X+Y)}} \ge \frac 1{\sqrt{J(X)}}+ \frac 1{\sqrt{J(Y)}},
 \ee
where, also in this case, equality holds if and only if both $X$ and $Y$ are Gaussian
random vectors.

Inequality \fer{sharp2} shows that, at least if applied to log-concave probability
densities, the functional $1/ \sqrt{J(\cdot)}$ behaves with respect to convolutions
like Shannon's entropy power \cite{Sha, Sta} and the reciprocal of Fisher information
\cite{Bla, Sta}. The fact that inequalities  \fer{entr}, \fer{BS} and \fer{sharp2}
share a common nature is further clarified by noticing that, when evaluated in
correspondence to the Gaussian vector $Z_\sigma$,
 \[
N(Z_\sigma)= \tilde I(Z_\sigma)= \sqrt{n/J(Z_\sigma)}= \sigma.
 \]
In addition to the present results, other inequalities related to Fisher information in one-dimension have been recently obtained in \cite{CG}. In  particular, the sign of the subsequent derivatives of Shannon's entropy $H(X+Z_t)$ up to order four have been computed explicitly. Since these derivatives alternate in sign, it is conjectured in \cite{CG} that this property has to hold for all subsequent derivatives. This is an old conjecture that goes back at least to McKean \cite{McK}, who investigated derivatives of Shannon's entropy up to the order three. Despite the title, however, in \cite{CG} the sign of the subsequent derivatives of the entropy power $N(X+Z_t)$ is not investigated.

The paper is organized as follows. In Section \ref{sec2}, we introduce the background and prove the concavity property of the reciprocal of Fisher information for log-concave densities. In Section \ref{sec3} we show that the third derivative in Costa's entropy power inequality, still when evaluated on log-concave densities, is non-negative. Last, Section \ref{sec4} will be devoted to the consequences of the new results on isoperimetric inequalities for entropies.

\section{The concavity property}\label{sec2}

We recall that a function $f$ on $\R^n$ is log-concave if it is of the
form
\be\label{log-c}
 f(x) = \exp\left\{-\Phi(x)\right\},
 \ee
for some convex function $\Phi: \R^n \to \R$. A prime example is the Gaussian density,
where $\Phi(x)$ is quadratic in $x$. Further, log-concave distributions include Gamma
distributions with shape parameter at least one, $Beta(\alpha, \beta)$ distributions
with $\alpha, \beta \ge1$, Weibull distributions with shape parameter at least one,
Gumbel, logistic and Laplace densities (see, for example, Marshall and Olkin
\cite{MO}). Log-concave functions have a number of properties that are desirable for
modelling.  Marginal distributions, convolutions and product measures of log-concave
distributions and densities are again log-concave (cf. for example, Dharmadhikari and
Joag-Dev \cite{DJ}).

Let $z_\sigma(x)$, with $x\in\R^n$, $n\ge1$, denote the density of the Gaussian random vector $Z_\sigma$ 
 \be\label{max}
z_\sigma(x) = \frac 1{(2\pi\sigma)^{n/2}}\exp \left\{- \frac{x^2}{2\sigma}\right\}.
 \ee
Assume that the random vector $X$ has a log-concave density $f(x)$, $x \in \R^n$.
Then, for any  $t>0$, the random vector $X+Z_t$, where the Gaussian $Z_t$ is
independent of $X$, has a density which is the convolution of the log-concave
functions $f$ and the (log-concave) Gaussian density $z_t$ defined in \fer{max}.
Therefore, for any $t>0$,  $X+Z_t$ has a log-concave smooth density function.  This
simple remark, coupled with the results of \cite{LT}, allows to justify the
computations that follows.

Let us evaluate the derivatives of $\tilde I(X+Z_t)$, with respect to $t$, $t>0$.
Thanks to \fer{JJ} we obtain
 \be\label{prima}
 \frac d{dt} \tilde I(X+Z_t) = n \,\frac{J(X+Z_t)}{I^2(X+Z_t)},
 \ee
and
\be\label{seconda}
 \frac {d^2}{dt^2} \tilde I(X+Z_t) = n\left( 2\, \frac{J^2(X+Z_t)}{I^3(X+Z_t)} - \frac{K(X+Z_t)}{I^2(X+Z_t)}\right).
 \ee
In \fer{seconda} we defined
 \be\label{KK}
K(X+Z_t) =  - \frac d{dt}I(X+Z_t).
 \ee
Hence, to prove concavity we need to show that, for log-concave densities
 \be\label{ineK}
K(X+Z_t) \ge  2 \frac{J^2(X+Z_t)}{I(X+Z_t)}.
 \ee
Note that
 \be\label{ide}
 I(Z_\sigma) = \frac n\sigma, \quad  J(Z_\sigma) = \frac n{\sigma^2}, \quad K(Z_\sigma) = 2 \frac n{\sigma^3}.
 \ee
Consequently, inequality \fer{ineK} is verified with the equality sign in
correspondence to a Gaussian random vector.

Using the second identity in \fer{ide} into \fer{sharp2} it is immediate to recover a
lower bound for $K(\cdot)$. This idea goes back to Dembo \cite{Dem},  who made
analogous use of the  Blachman--Stam inequality \fer{BS} to recover the sign of the
second derivative of the entropy power. Let $\sigma, t >0$. By choosing $X= W+
Z_\sigma$ and $Y = Z_t$, inequality \fer{sharp2} becomes
  \[
\frac 1{\sqrt{J(W+ Z_\sigma + Z_{t})}} \ge \frac 1{\sqrt{J(W + Z_\sigma)}}+ \frac t{\sqrt{n}}.
 \]
Then, for all $t>0$
 \[
 \frac 1t \left(\frac 1{\sqrt{J(W+ Z_\sigma + Z_{t})}} - \frac 1{\sqrt{J(W + Z_\sigma)}} \right)\ge \frac 1 {\sqrt{n}},
 \]
and this implies, passing to the limit $t \to 0^+$
 \[
\frac 12 \frac{K(W+Z_\sigma)}{J^{3/2}(W+Z_\sigma)} \ge \frac 1 {\sqrt{n}},
 \]
 for any $\sigma >0$. Hence, a direct application of inequality \fer{sharp2} shows that $K(X+Z_t)$ is bounded from below, and
  \be\label{ine-w}
K(X+Z_t) \ge  2 \frac{J^{3/2}(X+Z_t)}{\sqrt{n}}.
  \ee
Unfortunately, inequality \fer{ine-w} is weaker than \fer{ineK}, since it is known
that, for all random vectors $X$ and $Z_t$ independent from each other \cite{Dem, Vil}
 \be\label{dem}
{J(X+Z_t)}\ge \frac{I^2(X+Z_t)}n,
 \ee
and \fer{dem} implies
 \[
 \frac{J^2(X+Z_t)}{I(X+Z_t)} \ge \frac{J^{3/2}(X+Z_t)}{\sqrt{n}}.
 \]
To achieve the right result, we will work directly on inequality \fer{ine-main}. Let us fix $Y = Z_t$. Then, since for if $i \not=j$
 \[
 \int_{\R^n}  \frac{\partial_i z_t(x) \partial_j z_t(x) }{z_t(x)} \, dx = \int_{\R^n}  \frac{x_ix_j }{t^2} z_t (x) \, dx = 0,
 \]
one obtains
 \be\label{sempl}
 H(X,Z_t) = \sum_{i=1}^n \int_{\{f>0\}}\frac{f_i^2}f \, dx \int_{\R^n}  \frac{x_i^2 }{t^2} z_t \, dx = I(X)\frac 1nI(Z_t) = \frac 1t I(X).
 \ee
Hence, using \fer{ide} and \fer{sempl}, inequality \fer{ine-main} takes the form
 \be\label{ine-m}
 J(X+Z_t) \le \a^4 J(X) + (1-\a)^4\frac n{t^2} + 2\a^2(1-\a)^2 \frac 1t I(X).
 \ee
We observe that the function
 \[
 \Lambda(\a) = \a^4 J(X) + (1-\a)^4\frac n{t^2} + 2\a^2(1-\a)^2 \frac 1t I(X)
 \]
is convex in $a$, $0 \le \a \le1$. This fact follows by evaluating the sign of $\Lambda''(\a)$, where 
 \[
 \frac 1{12} \Lambda''(\a) = \a^2 J(X) + (\a-1)^2 \frac n{t^2} + \frac 13[(1-\a)^2 + 4\a(\a-1) +\a^2]\frac 1t I(X).
 \]
Clearly both $\Lambda''(0)$ and $\Lambda''(1)$ are strictly bigger than zero.  Hence,
$\Lambda(\a)$ is convex if, for $r = \a/(1-\a)$
 \[
 r^2 J(X) + \frac n{t^2} + \frac 13(r^2 -4r +1)\frac 1t I(X) \ge 0.
 \]
Now,
 \[
 \left( J(X) + \frac 1{3t} I(X) \right)r^2  - \frac 43\frac 1t I(X)r + \left( \frac n{t^2} + \frac 1{3t} I(X)
 \right) \ge
 \]
 \[
J(X) r^2 - 2\frac{I(X)}t r + \frac n{t^2} \ge 0.
 \]
The last inequality follows from \fer{dem}.

The previous computations show that, for any given value of $t>0$, there exists a
unique point $\bar\a = \bar\a(t)$ in which the function $ \Lambda(\a)$ attains the
minimum value. In correspondence to this optimal value,  inequality \fer{ine-m} takes
the equivalent optimal form
 \be\label{opt}
 J(X+Z_t) \le \bar\a(t)^4 J(X) + (1-\bar\a(t))^4\frac n{t^2} + 2\bar\a(t)^2(1-\bar\a(t))^2 \frac 1t I(X).
 \ee
The evaluation of $\bar\a(t)$ requires to solve a third order equation. However, since
we are interested in the value of the right-hand side of \fer{opt} for small values of
the variable $t$, it is enough to evaluate in an exact way the value of $\bar\a(t)$ up
to the order one in $t$. By substituting
 \[
\bar\a(t) = c_0 +c_1 t + o(t)
 \]
in the third order equation $\Lambda'(\a) = 0$, and equating the coefficients of $t$
at the orders $0$ and $1$ we obtain
 \be\label{val}
c_0 = 1, \quad c_1 = - \frac{J(x)}{I(X)}.
 \ee
Consequently, for $t \ll  1$
 \be\label{appr}
\Lambda(\bar\a(t) ) = J(X) -2\frac{J^2(X)}{I(X)}\, t +o(t).
 \ee
Finally, by using expression \fer{appr} into inequality \fer{opt} we obtain
 \begin{equations}\label{1or}
\frac 1{\sqrt{J(X+ Z_\sigma + Z_t)}} \ge \frac 1{\sqrt{J(X+Z_\sigma)
-2\frac{J^2(X+Z_\sigma)}{I(X+Z_\sigma)}\, t +o(t)}}= \\
\frac 1{\sqrt{J(X+Z_\sigma)}} + \frac{\sqrt{J(X+Z_\sigma)}}{I(X+Z_\sigma)}\, t +o(t),
 \end{equations}
which implies, for all $\sigma >0$, the inequality
 \be\label{fina}
\lim_{t\to0^+}\frac 1t \left( \frac 1{\sqrt{J(X+ Z_\sigma + Z_t)}} - \frac
1{\sqrt{J(X+Z_\sigma)}}\right) \ge \frac{\sqrt{J(X+Z_\sigma)}}{I(X+Z_\sigma)}.
 \ee
At this point, inequality \fer{ineK} follows from \fer{fina} simply by evaluating the
derivative of
 \[
J(X+Z_t)=\left(\frac 1{\sqrt{J(X+Z_t)}} \right)^{-2}.
 \]
This gives
 \begin{equations}\label{extra}
K(X+Z_t) = - \frac d{dt}J(X+Z_t) = -\frac d{dt}\left(\frac 1{\sqrt{J(X+Z_t)}}
\right)^{-2} =  \\
2\,J(X+Z_t)^{3/2}\frac d{dt}\frac 1{\sqrt{J(X+Z_t)}} \ge 2\,\frac{J^2(X+Z_t)}{I(X+Z_t)}.
\end{equations}
Hence we proved

\

\begin{thm}\label{Fi} Let $X$ be a random vector in $\R^n$, $n \ge 1$, such that its probability density $f(x)$ is log-concave. Then the reciprocal of the Fisher information of $X+Z_t$, where $X$ and $Z_t$ are independent each other, is concave in $t$, i.e.
 \[
\frac{d^2}{dt^2} \frac 1{I(X+Z_t)} \le 0.
 \]
\end{thm}

\section{An improvement of Costa's EPI}\label{sec3}

The computations of Section \ref{sec2} can be fruitfully used to improve Costa's result on concavity of the entropy power $N(X+Z_t)$.  To this aim, let us compute the derivatives in $t$ of $N(X+Z_t)$, up to the third order. The first derivative can be easily evaluated by resorting to de Bruijn identity
 \be\label{pri}
 \frac d{dt} H(X+Z_t) = \frac 12 I(X+Z_t).
 \ee
Then, identities \fer{JJ} and \fer{KK} can be applied to compute the subsequent ones. By setting $X+Z_t = W_t$ one obtains
 \be\label{uno}
 \frac d{dt} N(W_t) = \frac 1{n}N(W_t) I(W_t),
 \ee
and, respectively
 \be\label{due}
 \frac{d^2}{dt^2} N(W_t) = \frac 1{n}N(W_t)\left( \frac{I(W_t)^2}n -J(W_t) \right),
 \ee
 and
 \be\label{tre}
 \frac{d^3}{dt^3} N(W_t) = \frac 1{n} N(W_t)\left( K(W_t)  + \frac{I(W_t)^3}{n^2} -3\frac{I(W_t)J(W_t)}n \right).
 \ee
Note that, by virtue of identities \fer{dem} and \fer{ide}, the right-hand sides of both \fer{due} and \fer{tre} vanishes if $W_t$ is a Gaussian random vector. Using inequality \fer{ineK} we get
 \[
 K(W_t)  + \frac{I(W_t)^3}{n^2} -3\frac{I(W_t)J(W_t)}n \ge 2\frac{J^2(W_t)}{I(W_t)} + \frac{I(W_t)^3}{n^2} -3\frac{I(W_t)J(W_t)}n.
 \]
Thus, by setting $p = nJ(W_t)/I^2(W_t)$, the sign of the expression on the right-hand side of \fer{tre} will coincide with the sign of the expression
 \be\label{tre1}
 2p^2 -3p +1.
 \ee
Since $p \ge 1$ in view of the inequality \fer{dem} \cite{Dem, Vil},  $2p^2 -3p +1 \ge 0$, and the result follows. Last, the cases of equality coincide with the cases in which there is equality both in \fer{ineK} and \fer{dem}, namely if and only if $W_t$ is a Gaussian random vector.

\noindent We proved

\

\begin{thm}\label{EPI} Let $X$ be a random vector in $\R^n$, $n \ge 1$, such that its probability density $f(x)$ is log-concave. Then the entropy power of $X+Z_t$, where $X$ and $Z_t$ are independent each other, has the derivatives which alternate in sign up to the order three. In particular $N(X+Z_t)$ is concave in $t$, and
 \[
\frac{d^3}{dt^3} N(X+Z_t) \ge 0.
 \]
\end{thm}

\section{Isoperimetric inequalities}\label{sec4}

An interesting consequence of the concavity property of entropy power is the so-called \emph{isoperimetric inequality} for entropies \cite{Dem, DCT, Tos1}
\be\label{iso}
\frac 1n N(X)I(X) \ge 1.
\ee
which is easily obtained from \fer{uno}. Note that the quantity which is bounded from below coincides with the derivative of the entropy power of $X+Z_t$, evaluated as time $t \to 0$.  Indeed, the concavity of $N(X+Z_t)$ implies the non-increasing property of the right-hand side of \fer{uno} with respect to $t$, which, coupled with the scaling invariance of the product $N(f)I(f)$ with respect to dilation allows to identify the lower bound \cite{Tos1}.

Among others, the concavity property makes evident the connection between the solution to the heat equation and inequalities \cite{Tos1, Tos2, Tos3}. Resorting to this connection, the concavity property of entropy power has been recently shown to hold also for Renyi entropies \cite{ST, Tos5}.

Likewise, the concavity result of Theorem \ref{Fi} allows to recover an \emph{isoperimetric inequality} for Fisher information, that reads
 \be\label{isoF}
n \, \frac{J(X)}{I^2(X)} \ge 1.
 \ee
As in \fer{iso}, in \fer{isoF} the quantity which is bounded from below coincides with
the derivative of the reciprocal of the Fisher information of $X+Z_t$, evaluated as
time $t \to 0$. Inequality \fer{isoF}, which follows from \fer{dem} by letting $t \to 0$,  has been obtained  as a byproduct of Costa's
entropy power inequality \cite{Cos, Dem, Vil}, and holds for all random vectors with a
suitably smooth probability density. The novelty here is that, in the case in which
the probability density of $X$ is log-concave, the quantity $nJ(X+Z_t)/I^2(X+Z_t)$ is
non-increasing in time.

However, a new inequality is obtained for random vectors with log-concave densities. In this case in fact, by taking the limit $t \to 0$ in \fer{ineK} we get
  \be\label{isoK}
K(X) \ge 2\, \frac{J^2(X)}{I(X)}.
 \ee
In one-dimension, the expression of $K(X+Z_t)$ has been evaluated in \cite{McK}. Then,
a more convenient expression which allows to recover its positivity has been obtained recently
in \cite{CG}. In general, written as
 \[
J^2(X) \le \frac 12 I(X)K(X)
 \]
inequality \fer{isoK} provides a sharp upper bound on $J^2(X)$, which is an expression
containing second derivatives of the logarithm of the log-concave density $f(x)$, in
terms of the product of the Fisher information, which depends on first derivatives of
the logarithm, and $K(X)$, which depends on derivatives of the logarithm up to order
three. Unfortunately, these expressions are heavy and difficult to handle.

\section{Conclusions}

As recently shown in \cite{Tos4},  log-concave densities exhibit a number of
interesting properties. In this paper, we proved two new properties, both related to
the behavior in time of various functionals evaluated on the sum of a random vector
$X$ with a log-concave probability density with an independent Gaussian vector $Z_t=
N(0,tI_n)$. First, we proved that the reciprocal of Fisher information of $X+N_t$ is
concave in $t$, thus extending Costa's result on Shannon's entropy power to
Blachman--Stam inequality. Second, we showed that Shannon's entropy power of $X+Z_t$
has derivatives in $t$ that alternate in sign up to order three. In addition to
Costa's concavity property of entropy power, which concerns the sign of the
second-order derivative, it has been here discovered that also the third derivative of
Shannon's entropy power has a (universal) sign.

\bigskip \noindent

{\bf Acknowledgment:} This work has been written within the activities of the National
Group of Mathematical Physics of INDAM (National Institute of High Mathematics). The
support of the  project ``Optimal mass transportation, geometrical and functional
inequalities with applications'', financed by the Minister of University and Research,
is kindly acknowledged.

\end{document}